# Computational studies on magnetism and ferroelectricity

Ke Xu(徐可)[1], Junsheng Feng(冯俊生)[2], and Hongjun Xiang(向红军)[3,†]

[1]*Hubei Key Laboratory of Low Dimensional Optoelectronic Materials and Devices, School of Physics and Electronics Engineering, Hubei University of Arts and Science, Xiangyang 441053, P. R. China*

[2]*School of Physics and Material Engineering, Hefei Normal University, Hefei 230601, P. R. China*

[3]*Key Laboratory of Computational Physical Sciences (Ministry of Education), State Key Laboratory of Surface Physics, and Department of Physics, Fudan University, Shanghai 200433, P. R. China*

[†]Corresponding author. E-mail: hxiang@fudan.edu.cn

**Abstract**

Magnetics, ferroelectrics and multiferroics have attracted great attentions because they are not only extremely important for investigating fundamental physics, but also have important applications in information technology. Here, recent computational studies on magnetism and ferroelectricity are reviewed. We first give a brief introduction to magnets, ferroelectrics, and multiferroics. Then, theoretical models and corresponding computational methods for investigating these materials are presented. In particular, a new method for computing the linear magnetoelectric coupling tensor without applying an external field in the first principle calculations is proposed for the first time. The functionalities of our home-made Property Analysis and Simulation Package for materials (PASP) and its applications in the field of magnetism and ferroelectricity are discussed. Finally, we summarize this review and give a perspective on possible directions of future computational studies on magnetism and ferroelectricity.

# 1. Introduction

Magnetic materials are widely used in a diverse range of applications, including information storage, electronics, and even in biomedicine. [1] The collinear magnetic materials (e.g., ferromagnet, antiferromagnet) have been studied for a long time. However, the exotic magnetic states attract a lot of interest recently in both theoretical and experimental studies, such as spin glasses, [2] spin ice, [3] spin liquid, [4,5] skyrmions [6,7] and other topological spin textures [8]. In 2017, the ferromagnetism in monolayer $Cr_2Ge_2Te_6$ [9] and $CrI_3$ [10] were reported experimentally. After that, the two-dimensional (2D) magnetic materials has received considerable attention, and several related studies have been carried out on the few layers $CrI_3$ and $Cr_2Ge_2Te_6$. [11–14] Other 2D FM materials were also predicted theoretically or synthesized experimentally, such as $Fe_3GeTe_2$, [15] $VSe_2$, [16] $MnSe_2$, [17] $Fe_3P$, [18], $VI_3$ [19] and van der Waals (vdW) structures. [20]

Similar to magnetism, ferroelectricity is also an appealing property because it can be utilized to make functional devices, especially electronic devices such as nonvolatile memory, tunable capacitor, solar cell and tunnel junction [21]. Ferroelectric (FE) materials are a large family, including inorganic oxides, [22] organic-inorganic hybrid materials, [23] organic compounds, [24] liquid crystals, [25] and polymers. [26] In the past decade, with the rapid development of thin film technologies, 2D FEs have attracted much attention. Ferroelectricity in 2D materials with stable layered structures and reduced surface energy may be stable even in the presence of a depolarization field, thereby opening a pathway to explore low dimensional ferroelectricity. So far, various 2D van der Waals (vdW) materials with the thickness of few layers or even monolayer have been reported to be intrinsically FE, such as $MX_2$ (M = W Mo, X = S, Se, Te) transition metal dichalcogenides (TMDs), [27] group-IV monochalcogenides, [28] metal triphosphates, [29] layer perovskites, [30] indium selenide ($In_2Se_3$), [31] and sliding FEs. [32]

Magnetism and ferroelectricity are like the front and back of a coin, which can be combined in a single material dubbed as multiferroic. Multiferroic materials may be ideal candidates for applications in novel multifunctional magnetoelectric (ME) devices and high performance information storage and processing devices. [33] Two breakthrough studies in 2003 bring up the renaissance of multiferroics. In one study, the epitaxial $BiFeO_3$ thin film is successfully prepared and demonstrated the room-temperature multiferroic behavior. [34] In another study, Tokura and coworkers discovered the colossal ME effect in TbMnO$_3$. [35] Then multiferroics are classified into two types (i.e., type-I and type-II) according to the different mechanisms of ferroelectric polarization by Khomskii. [36] In fact, the ME effect in type-I multiferroics (e.g., BiFeO$_3$) [34] is usually weak, since the ferroelectricity and magnetism are independent of each other. While in type-II multiferroics (e.g., TbMnO$_3$), [35] the ME effect is expected to be much stronger, because the ferroelectricity is induced directly by a special spin order. Very recently, the 2D ferroelastic-ferroelectric, [37] and 2D ferroelectric-ferromagnetic materials [38–40] were also reported. Recently, Song *et al.* reported [41] the first experimental discovery of type-II multiferroic order in a 2D vdW material NiI$_2$.

In the past decades, first principle based studies have made significant contributions to the field of magnetism and ferroelectricity. Density functional theory (DFT) calculations have been adopted to investigating various topics including the magnetic domain wall induced electric polarization in Gd$_3$Fe$_5$O$_{12}$, [42] skyrmions in Ni-halide monolayer, [43] Berezinskii-Kosterlitz-Thouless (BKT) phase existing between the FE and paraelectric (PE) states in BaTiO$_3$ ultrathin film [44] and single layer SnTe, [45] the proximate quantum spin liquid states in X[Pd(dmit)$_2$]$_2$, [46] frustrated magnetism in Mott insulating (V$_{1-x}$Cr$_x$)$_2$O$_3$ [47] and topological surface states in superlatticelike MnBi$_2$Te$_4$/(Bi$_2$Te$_3$)$_n$ [48]. In addition, first principle based effective Hamiltonian models were now widely adopted to investigate the ground state at zero temperature and temperature dependent properties. [44,45,49–52] For example, with the effective Hamiltonian model approach, Bellaiche *et al.* predicted a magnetic phase diagram in BiFeO$_3$ as a function of first- and second-nearest-neighbor interaction strength and find new spin arrangement. [49] The effective Hamiltonian method was also used to investigate the FE-antiferrodistortive coupling in small tolerance factor perovskite. [51]

This remaining sections of this review are organized as follows. In sec. 2, we will briefly introduce the microscopic models about magnetism, ferroelectricity, multiferroicity and how to calculate ME coupling tensor $\alpha_{ij}$. Next, in sec. 3 we will introduce the four-state method to compute the various parameters and our Property Analysis and Simulation Package for materials (PASP). In sec. 4, we will demonstrate the applications of the models, methods and

PASP to various magnetic and ferroelectric systems. [51] [60] [73] At last, we summarize the review and give a perspective on future computational developments in the field of magnetism and ferroelectricity.

## 2. Theoretical models

Although DFT played an important role on studying magnetic and ferroelectric materials, it has some limitations: it is computationally demanding and thus is not suitable for large-scale systems; it has difficult to consider effects of temperature and external field; it gives the total results, but does not directly provide microscopic insight. Therefore, many theoretical models were developed to address these issues. In particular, effective Hamiltonian models were proposed to understand the microscopic mechanisms of the magnetism, ferroelectricity, and multiferroicity, and to simulate the behaviors of ferroic materials. In this section, we will first discuss the effective spin Hamiltonian models for magnetic materials. Then, we briefly introduce the effective Hamiltonian model for ferroelectrics, which provides an efficient way for computing the thermodynamic properties and simulating the dynamics behaviors of ferroelectrics. Finally, we will discuss two theoretical models related to multiferroics.

### 2.1 Spin Hamiltonian models

The effective magnetic Hamiltonian can be written as:

$$\begin{aligned}
\mathcal{H}_{spin} &= \mathcal{H}_{ex} + \mathcal{H}_{SIA} + \mathcal{H}_{higher} \\
&= \sum_{\langle ij \rangle} \boldsymbol{S}_i^T \cdot \boldsymbol{\mathcal{J}}_{ij} \cdot \boldsymbol{S}_j + \sum_i \boldsymbol{S}_i^T \cdot \boldsymbol{\mathcal{A}}_{ii} \cdot \boldsymbol{S}_i \\
&+ \sum_{\langle ij \rangle} K_{ij} (\boldsymbol{S}_i \cdot \boldsymbol{S}_j)^2 + \sum_{\langle ijk \rangle} K_{ijk} (\boldsymbol{S}_i \cdot \boldsymbol{S}_j)(\boldsymbol{S}_j \cdot \boldsymbol{S}_k) \\
&+ \sum_{\langle ijkl \rangle} K_{ijkl} \left[ (\boldsymbol{S}_i \cdot \boldsymbol{S}_j)(\boldsymbol{S}_k \cdot \boldsymbol{S}_l) + (\boldsymbol{S}_i \cdot \boldsymbol{S}_l)(\boldsymbol{S}_j \cdot \boldsymbol{S}_k) - (\boldsymbol{S}_i \cdot \boldsymbol{S}_k)(\boldsymbol{S}_j \cdot \boldsymbol{S}_l) \right] \\
&+ \cdots \cdots
\end{aligned} \quad (1)$$

Eq. (1) includes the bilinear interactions $\mathcal{H}_{ex}$ and $\mathcal{H}_{SIA}$, and the higher order interactions $\mathcal{H}_{higher}$. The summation indices <$ijk$> and <$ijkl$> mean that the sum goes over triangles of sites and the sum is take over all distinct four-spin cycles on square plaquettes, respectively. For the bilinear interactions, the Heisenberg exchange parameters $J_{ij}$, the Dzyaloshinskii–Moriya (DM) interaction [53] vector $\boldsymbol{D}_{ij}$ and the Kitaev interaction parameters $K_\gamma$ can be extracted from the full $\mathcal{J}$ matrix (as will be defined in 2.1.2). The single ion anisropy (SIA) Hamiltonian is characterized by the $\mathcal{A}$ matrix. The calculation of these magnetic parameters will be discussed in section 3.1.

### 2.1.1 Heisenberg model

By applying the perturbation approach to the Hubbard model, [54] the effective spin Hamiltonians such as celebrated simplest or isotropic Heisenberg model can be obtained by down-folding the fermionic degrees of freedom into the proper low energy subspaces. [54] The Heisenberg model can be written as:

$$\mathcal{H}_{spin} = \sum_{\langle ij \rangle} J_{ij} \boldsymbol{S}_i \cdot \boldsymbol{S}_j \qquad (2)$$

For magnetic systems, the Heisenberg spin exchange constants in a given system can be calculated by the four-state energy mapping method within the DFT framework as discussed below. [55,56] When $J_{ij} > 0$ and $J_{ij} < 0$, antiferromagnetic (AFM) and ferromagnetic (FM) configuration are preferred.

### 2.1.2 Kitaev model

The exactly solvable Kitaev model [57] is a bond-dependent interaction of spin-$\frac{1}{2}$ model on a 2D honeycomb lattice, in which the spins fractionalize into Majorana fermions and form a topological quantum spin liquid (QSL) in the ground state. $\alpha$-RuCl$_3$ [58] is a 2D magnetic insulators with a honeycomb structure, and it was noticed that they accommodate essential ingredients of the Kitaev model owing to the interplay of electron correlation and spin–orbit coupling. The Kitaev model consists of spin-$\frac{1}{2}$ on a honeycomb lattice, with nearest-neighbour Ising interactions with bond-dependent easy axes parallel to the *x*, *y* and *z* axes. The orthogonal anisotropies of the three nearest-neighbour bonds of each spin conflict with each other, leading to the spin frustration. [59] In 2018, Xu *et al.* predicted the existence of Kitaev interactions in 2D CrI$_3$ and CrGeTe$_3$-like systems. [60] In 2019, Hae-Young Kee *et al.* present a theory of the spin-1 Kitaev interaction in two-dimensional edge-shared octahedral systems. [61] It was recently proposed that spin-$\frac{3}{2}$ QSL can be realized in 3d transition metal compound CrSiTe$_3$. [62]

The Kitaev interaction originate spin-orbital coupling (SOC) effect, and its terms can be derived from the $\mathcal{J}$ matrix. The $\mathcal{J}$ matrix characterizing the magnetic exchange couplings are expressed in the most general 3×3 matrix form as:

$$\mathcal{J} = \begin{pmatrix} J_{xx} & J_{xy} & J_{xz} \\ J_{yx} & J_{yy} & J_{yz} \\ J_{zx} & J_{zy} & J_{zz} \end{pmatrix} = J^H I_{3\times 3} + \left[ \frac{\mathcal{J} + \mathcal{J}^T}{2} - J^H I_{3\times 3} \right] + \frac{\mathcal{J} - \mathcal{J}^T}{2} \qquad (3)$$

The $\mathcal{J}$ matrix can be decomposed into three parts: the first two terms are symmetric exchanges, while the last term is antisymmetric exchange. The first term is the Heisenberg exchange, where $J^H$ is defined as $J^H = \frac{1}{3}(J_{xx} + J_{yy} + J_{zz})$; the second term is symmetric anisotropic exchange $K_{ij} = \frac{1}{2}(J_{ij} + J_{ji})$ ($i \neq j$) and the Kitaev interaction $K_{ii} = J_{ii} - J^H$; and the last term is the so-called DM interaction with $D_{ij} = \frac{1}{2}(J_{ij} - J_{ji})$.

### 2.1.3 Higher order interactions

Beyond the bilinear interactions, higher order exchange interactions involving the hopping of two or more electrons sometimes play a crucial role in exotic magnetic materials, [63] especially if the magnetic atoms have large magnetic moments or if the system is itinerant. [1] Applying the fourth order perturbation theory with the Löwdin's downfolding technique to the multiorbital Hubbard model, the fourth order interaction can be derived. [54] The fourth order interactions consist two-body fourth-order interaction $K_{ij}(\boldsymbol{S}_i \cdot \boldsymbol{S}_j)^2$ [i.e., biquadratic exchange (BQ) interaction], three-body fourth-order interaction $K_{ijk}(\boldsymbol{S}_i \cdot \boldsymbol{S}_j)(\boldsymbol{S}_j \cdot \boldsymbol{S}_k)$ and four-body fourth-order interaction $K_{ijkl}(\boldsymbol{S}_i \cdot \boldsymbol{S}_j)(\boldsymbol{S}_k \cdot \boldsymbol{S}_l)$ [see Eq. (1)].

Higher order interactions are found to be crucial to the magnetic features in several systems, such as multilayer materials, [64] perovskites, [65,66] and 2D magnets. [67,68] Orthorhombic perovskite o-HoMnO$_3$ [65] is reported to be E-type antiferromagnetic (E-AFM) in the experiment. A theoretical study suggested that the E-AFM state can be properly predicted only when the BQ exchange term is included in the model Hamiltonian. [66] In 2D vdW magnets, the higher order magnetic interactions were also found to play a key role on the magnetic properties. [63] In the single layer NiCl$_2$, the nearest-neighbor BQ interaction is necessary for FM ground state prediction. [67] Furthermore, the four site four spin interaction term was found to have a large effect on the energy barrier preventing skyrmion and antiskyrmion collapse into the FM state in Pd/Fe/Rh(111) ultrathin film. [69]

## 2.2 Effective Hamiltonian models for ferroelectrics

Following the approach of Vanderbilt and his coworkers, [70] the effective Hamiltonian is written in terms of the FE degree of freedom $\{u\}$ and the homogeneous strains $\{\eta_l\}$ as:

$$\mathcal{H}_{eff} = E^{self}(\{u\}) + E^{dpl}(\{u\}) + E^{short}(\{u\}) + E^{elas}(\{\eta_l\}) + E^{int}(\{u\},\{\eta_l\}) \tag{4}$$

where the effective Hamiltonian contains five parts: (i) a local-mode self-energy $E^{self}(\{u\}) = \sum_i E(u_i)$, where $E(u_i)$ is the energy of an isolated local mode at cell $R_i$ with amplitude $u_i$ with respect to the PE structure; (ii) long-range dipole-dipole interaction, $E^{dpl}(\{u\}) = \frac{Z^{*2}}{\varepsilon_\infty}\sum_{i<j}\frac{u_i \cdot u_j - 3(\hat{R}_{ij} \cdot u_i)(\hat{R}_{ij} \cdot u_j)}{R_{ij}^3}$, where $Z^*$ is the Born effective charge and $\varepsilon_\infty$ is the electronic dielectric constant of the FEs; (iii) $E^{short}(\{u\})$ is the energy contribution due to the short range interactions between neighboring local modes, with dipole-dipole interactions excluded. Together with the dipole-dipole interaction, this interaction determines the soft-mode energy away from the zone center. Expanded up to second order of $\{u\}$, it can be written as: $E^{short}(\{u\}) = \sum_{i \neq j}\sum_{\alpha\beta} J_{ij,\alpha\beta} u_{i\alpha} u_{j\beta}$. The coupling matrix $J_{ij,\alpha\beta}$ is a function of $R_{ij}$ and should decayed rapidly with increasing $R_{ij}$; (iv) The strain energy $E^{elas}(\{u\})$ simply depends on the homogeneous strain and inhomogeneous strain characterized by the local strain variables $\eta_l(R_i)$ (the Voigt notation is used with $l = 1 \sim 6$). It is included to describe the elastic deformation of the system; (v) To describe the coupling between the elastic deformations and the local modes, the on-site interaction is adopted: $E^{int}(\{u\},\{\eta_l\}) = \frac{1}{2}\sum_i \sum_{l\alpha\beta} B_{l\alpha\beta}\, \eta_l(R_i) u_{i\alpha}(R_i) u_{i\beta}(R_i)$. [70]

Parameters in such an effective Hamiltonian can be fitted by the energy mapping method based on DFT calculations with numerous configurations. The effective Hamiltonian method combined with the MC simulations can not only predict FE phase transitions from the temperature dependence of <*u*> and <*η*>, [48,50] but also construct the temperature-pressure phase diagram. [71]

## 2.3 Theoretical models for multiferroics

As we mentioned above, there exist both magnetism and ferroelectricity in a single multiferroic system. The coupling between magnetism and ferroelectricity is particularly intriguing as it not only is of great scientific importance, but also provides the basis for making novel energy-saving memory devices. Here, we discuss two theoretical models related to the ME coupling. In the first model, we describe a unified model for spin order induced improper ferroelectric polarization, which can explain the polarization induced by any spin structures: collinear, cycloidal spiral, proper screw, etc. [72−76] In the second model, we provide a new model for investigating the linear ME coupling effect [i.e., the first order change of magnetization (or electric polarization) response to the external electric (or magnetic) field].

### 2.3.1 Unified polarization model for spin-order induced ferroelectricity

For type-II multiferroics, the total electric polarization $\boldsymbol{P}_t$ can be viewed as a function of the spin directions $\boldsymbol{S}_i$ of the magnetic ions $i$, the displacements $\boldsymbol{u}_k$ of the ions $k$ and homogeneous strain $\eta_l$: [73−76]

$$\boldsymbol{P}_t = \boldsymbol{P}_t\left(\boldsymbol{S}_1, \boldsymbol{S}_2, \cdots, \boldsymbol{S}_m; \boldsymbol{u}_1, \boldsymbol{u}_2, \cdots, \boldsymbol{u}_n; \eta_1, \eta_2, \cdots, \eta_6\right) \tag{5}$$

here the ion displacements $\boldsymbol{u}_k$ induced by spin order and homogeneous strain $\eta_l$ are given with respect to a PE reference structure. For simplicity, we use the notation $\boldsymbol{u} = (\boldsymbol{u}_1, \boldsymbol{u}_2, \ldots, \boldsymbol{u}_n)$ and $\eta = \left(\eta_1, \eta_2, \cdots, \eta_6\right)$. Owing to the ion displacements induced by spin order and homogenous strain are both rather small, the total FE polarization is simplified to be:

$$\boldsymbol{P}_t \approx \boldsymbol{P}_e(\boldsymbol{S}_1, \boldsymbol{S}_2, \ldots, \boldsymbol{S}_m; \boldsymbol{u} = 0; \eta = 0) + \boldsymbol{P}_{ion,lattice}(\boldsymbol{u}; \eta) \tag{6}$$

the pure electronic contribution $P_e$ arises from the electron density redistribution induced by the spin order, and the lattice deformation contribution $P_{ion,lattice}$ originates from the spin order induced ion displacements and stress.

To obtain the pure electronic contribution $P_e$, let us consider a spin dimer with spatial inversion symmetry at the center, the distance vector from spin 1 to spin 2 will be taken along the $x$ axis. When a noncollinear spin configuration is applied on the dimer, the inversion symmetry will be broken, hence ferroelectric polarization $P$ will be induced. In general, the polarization $P$ is a function of the directions of spin 1 and spin 2, i.e, $P = P(S_{1x}, S_{1y}, S_{1z}; S_{2x}, S_{2y}, S_{2z})$. The $P$ can be expanded as a Taylor series of $S_{i\alpha}$ ($i = 1, 2$; $\alpha = x, y, z$). The odd terms of the Taylor expansion should vanish due to the time reversal symmetry. To neglect the fourth and higher order terms, the $P$ is written as:

$$P = P_1(S_1) + P_2(S_2) + P_{12}(S_1, S_2) \tag{7}$$

where the intrasite polarization $P_i(S_i)$ ($i = 1, 2$) is given by:

$$P_i(S_i) = \sum_{\alpha\beta} P_i^{\alpha\beta} S_{i\alpha} S_{i\beta} \tag{8}$$

here the expansion coefficients $P_i^{\alpha\beta}$ ($i = 1, 2$) and $P_{12}^{\alpha\beta}$ are vectors. The intersite term $P_{12}$ can be written as:

$$P_{12}(S_1, S_2) = \sum_{\alpha\beta} P_{12}^{\alpha\beta} S_{1\alpha} S_{2\beta} = \begin{pmatrix} S_{1x} & S_{1y} & S_{1z} \end{pmatrix} \begin{pmatrix} P_{12}^{xx} & P_{12}^{xy} & P_{12}^{xz} \\ P_{12}^{yx} & P_{12}^{yy} & P_{12}^{yz} \\ P_{12}^{zx} & P_{12}^{zy} & P_{12}^{zz} \end{pmatrix} \begin{pmatrix} S_{2x} \\ S_{2y} \\ S_{2z} \end{pmatrix} = S_1^T P_{int} S_2 \tag{9}$$

where $P_{int}$ is a matrix in which each element is a vector. For easily understanding, the $P_{int}$ can be decomposed into the isotropic symmetric diagonal matrix $P_J$, antisymmetric matrix $P_D$, and anisotropic symmetric matrix $P_\Gamma$:

$$P_J = \frac{1}{3} \begin{pmatrix} P_{12}^{xx} + P_{12}^{yy} + P_{12}^{zz} & 0 & 0 \\ 0 & P_{12}^{xx} + P_{12}^{yy} + P_{12}^{zz} & 0 \\ 0 & 0 & P_{12}^{xx} + P_{12}^{yy} + P_{12}^{zz} \end{pmatrix} \tag{10}$$

$$P_D = \frac{1}{2}\begin{pmatrix} 0 & P_{12}^{xy} - P_{12}^{yx} & P_{12}^{xz} - P_{12}^{zx} \\ P_{12}^{yx} - P_{12}^{xy} & 0 & P_{12}^{yz} - P_{12}^{zy} \\ P_{12}^{zx} - P_{12}^{xz} & P_{12}^{zy} - P_{12}^{yz} & 0 \end{pmatrix} \quad (11)$$

$$P_\Gamma = \begin{pmatrix} P_{12}^{xx} - \frac{1}{3}(P_{12}^{xx} + P_{12}^{yy} + P_{12}^{zz}) & \frac{1}{2}(P_{12}^{xy} + P_{12}^{yx}) & \frac{1}{2}(P_{12}^{xz} + P_{12}^{zx}) \\ \frac{1}{2}(P_{12}^{yx} + P_{12}^{xy}) & P_{12}^{yy} - \frac{1}{3}(P_{12}^{xx} + P_{12}^{yy} + P_{12}^{zz}) & \frac{1}{2}(P_{12}^{yz} + P_{12}^{zy}) \\ \frac{1}{2}(P_{12}^{zx} + P_{12}^{xz}) & \frac{1}{2}(P_{12}^{zy} + P_{12}^{yz}) & P_{12}^{zz} - \frac{1}{3}(P_{12}^{xx} + P_{12}^{yy} + P_{12}^{zz}) \end{pmatrix} \quad (12)$$

The general formula for the polarization of a spin dimer can be reduced in certain cases. When the SOC effect is not included, it can be written as $P_{12}^{xx} = P_{12}^{yy} = P_{12}^{zz}$, corresponding to the usual exchange striction model. [73] If the spin dimer contains the inversion symmetry, the intersite polarization can be simplified as $P_{12} = M(S_1 \times S_2)$ with the 3×3 matrix $M$:

$$M = \begin{pmatrix} (P_{12}^{yz})_x & (P_{12}^{zx})_x & (P_{12}^{xy})_x \\ (P_{12}^{yz})_x & (P_{12}^{zx})_y & (P_{12}^{xy})_y \\ (P_{12}^{yz})_x & (P_{12}^{zx})_z & (P_{12}^{xy})_z \end{pmatrix} \quad (13)$$

Note that the KNB model [72] is a special case of the intersite polarization with $(P_{12}^{zx})_z = (P_{12}^{xy})_y$. The intersite term $M(S_1 \times S_2)$ can be referred as the generalized KNB (gKNB) model. The bond polarization model is a special case of the intrasite term $P_i(S_i)$. For instance, in the cubic perovskite LaMn$_3$Cr$_4$O$_{12}$, spin-driven ferroelectricity is demonstrated to be caused by the anisotropic symmetric exchange mechanism (i.e., the $P_\Gamma$ term). [77] Although the anisotropic symmetric exchange is usually weak, it becomes even stronger than the symmetric exchange striction term and spin-current term for a nonlinear M-O-M cluster in the large SOC limit.

Besides the spin order induced atomic displacements, the spin order induced lattice strain may give rise to an additional electric polarization through the piezoelectric effect. [78,79] In general, the total energy of a magnetic system can be written as $E(u_m, \eta_j, S_i) = E_{PM}(u_m, \eta_j) + E_{spin}(u_m, \eta_j, S_i)$, where $u_m$ is the ion displacement of reference structure for all the ions, $\eta_j$ is the homogeneous strain in Voigt notation, and $S_i$ is the spin vector of all the magnetic ions. With neglecting the terms of higher order, the Taylor series expansion of the paramagnetic state energy $E_{PM}(u_m, \eta_j)$ is:

$$E_{PM} = E_0 + \sum_m A_m u_m + \sum_j A_j \eta_j + \frac{1}{2}\sum_{mn} B_{mn} u_m u_n + \sum_{jk} B_{jk} \eta_j \eta_k + \sum_{mj} B_{mj} u_m \eta_j \qquad (14)$$

The first order coefficients $A_m$ and $A_j$ refer to the force and stress, respectively. The second order coefficients $B_{mn}$, $B_{jk}$ and $B_{mj}$ represent the force constant, frozen-ion elastic constant, and internal displacement tensor, respectively. The reference structure is in equilibrium in the PM state: $\frac{\partial E_{PM}}{\partial u_m} = \frac{\partial E_{PM}}{\partial \eta_l} = 0$ with $A_m = A_j = 0$.

The ion displacement and lattice deformation caused by the spin order can be obtained by minimizing the total energy $E(\boldsymbol{u}_m, \eta_j, \boldsymbol{S}_i)$ with respect to $u_m$ and $\eta_j$, that is: $\frac{\partial E(u_m,\eta_j,S_i)}{\partial u_m} = 0$ and $\frac{\partial E(u_m,\eta_j,S_i)}{\partial \eta_j} = 0$. If the system in the PM state is piezoelectric, the lattice deformation induced by spin order may give rise to an additional electric polarization. The polarization induced by the ion displacement $u_m$ and strain $\eta_j$ can also be computed with $P_\alpha = Z_{\alpha m} u_m + e_{\alpha j}\eta_j$, where $Z_{\alpha m}$ and $e_{\alpha j}$ are the Born effective charge and frozen-ion piezoelectric tensor, respectively. This unified model can not only describe the spin-lattice coupling in 2D CrI$_3$, CrGeTe$_3$ [80] and monoxide NiO, [81] but also explain the ferroelectricity of the 2D multiferroic NiI$_2$. [41]

### 2.3.2 A model for linear magnetoelectric coupling

The linear ME effect refers to a linear response of electric polarization $\Delta\boldsymbol{P}$ to an applied magnetic field $\boldsymbol{H}$, or the magnetization $\Delta\boldsymbol{M}$ induced by an electric field $\boldsymbol{E}$, i.e. $\Delta P_i = \alpha_{ij} H_j$ and $\Delta M_j = \alpha_{ij} E_i$, where $\alpha$ is the ME tensor. [82] The underlying mechanism of the linear ME effect comes from the breaking of spatial inversion symmetry (I) and time reversal symmetry (T). Note that a linear ME material (e.g., Cr$_2$O$_3$) may display the joint symmetry TI of spatial inversion symmetry and time reversal symmetry. The linear ME effect are found in Cr$_2$O$_3$, [83] FeS, [84] NaMnF$_3$, [85] BiFeO$_3$, [86] double layered CrI$_3$ [5] and perovskite superlattice. [87] The ME coupling tensor consists of two parts: (i) a "clamped-ion" contribution that accounts for ME effects occurring in an applied field with all ionic degrees of freedom remaining frozen, which is label $\alpha^{el}$; (ii) lattice relaxation in response to the external field, named $\alpha^{latt}$. The total response tensor is $\alpha^{tot} = \alpha^{el} + \alpha^{latt}$. Although the linear ME tensor has 9 components, the non-zero components depend on the magnetic symmetry of the system. [83−87] The methodology to calculate ME tensor was implemented by modifying the code in VASP package through calculating the electric polarization induced by an applied Zeeman $H$ field which only couples to the spin component of the magnetization. [83] Here, the response of electron spin is obtained by "clamping" the ions during the calculation; the relaxation of ionic positions in response to the $H$ field yields the sum of the ionic and electron spin components. The magnetic field is applied self-consistently and the calculations are done with setting a non-collinear spin configuration and including SOC effect. Actually, it is very challenging to calculate the ME coupling coefficient by applied a magnetic field. Because the forces induced by the applied magnetic field on the ions are rather small, the very high convergence criterion of the ionic forces is needed. In order to obtain the accurate ME response, the ionic forces need to be reduced to less than 5 $\mu$eV·Å$^{-1}$ [87] in practical calculations.

Usually, an external field ($\varepsilon$ or $H$) needs to be applied when calculating the ME coupling coefficient. In a pioneering study by Spaldin and coworkers, the polarization was calculated self-consistently by adding a Zeeman magnetic field to the DFT Kohn-Sham potential. [83] The magnetic field will change the spin orientations of magnetic ions, resulting in a change in the electric polarization due to the ME coupling. However, the calculation of the ME response by applying a magnetic field in first-principles calculations requires extremely high precision and this is very time-consuming. Recently, Chai *et al.* employed the unified polarization model to the case in the presence of an external magnetic field. [88] By performing magnetic symmetry analysis and calculating the spin susceptibility tensor $\chi_m$, linear and quadratic ME effects can be obtained. However, a first-principles scheme based on this method is still lacking. In this review, we propose for the first time a new method based on DFT calculations for computing the ME coupling parameters without applying an external field in the DFT calculations. First, we compute the ionic displacement and strain induced by a small external electric field. Then we can estimate the change of the exchange interactions induced by the effect of the external electric field. Finally, we can estimate the change of magnetization by finding the magnetic ground state of the system under the external electric field.

The total energy $E$ is a function of electric field $\varepsilon$, ionic displacement $\boldsymbol{u}$, and homogenous strain $\eta$ with respect to a reference system ($\varepsilon = \boldsymbol{u} = \eta = 0$), and it can be expanded up to second order with respect to ($\varepsilon$, $\boldsymbol{u}$, $\eta$):

$$\begin{aligned}
E(\boldsymbol{\varepsilon},\boldsymbol{u},\eta) &= E_0 + \sum_\alpha \frac{\partial E}{\partial \varepsilon_\alpha}\varepsilon_\alpha + \sum_{i\alpha}\frac{\partial E}{\partial u_{i\alpha}}u_{i\alpha} + \sum_j \frac{\partial E}{\partial \eta_j}\eta_j \\
&\quad + \frac{1}{2}\sum_{ij\alpha\beta}\frac{\partial^2 E}{\partial u_{i\alpha}\partial u_{j\beta}}u_{i\alpha}u_{j\beta} + \frac{1}{2}\sum_{jk}\frac{\partial^2 E}{\partial \eta_j \partial \eta_k}\eta_j\eta_k + \frac{1}{2}\sum_{\alpha\beta}\frac{\partial^2 E}{\partial \varepsilon_\alpha \partial \varepsilon_\beta}\varepsilon_\alpha\varepsilon_\beta \\
&\quad + \sum_{i\alpha j}\frac{\partial^2 E}{\partial u_{i\alpha}\partial \eta_j}u_{i\alpha}\eta_j + \sum_{i\alpha}\frac{\partial^2 E}{\partial \eta_i \partial \varepsilon_\alpha}\eta_i\varepsilon_\alpha + \sum_{i\alpha\beta}\frac{\partial^2 E}{\partial u_{i\alpha}\partial \varepsilon_\beta}u_{i\alpha}\varepsilon_\beta \\
&= E_0 - \boldsymbol{P}\cdot\boldsymbol{\varepsilon} + \frac{1}{2}\sum_{ij\alpha\beta}\Phi_{ij}^{\alpha\beta}u_{i\alpha}u_{j\beta} + \frac{1}{2}\sum_{jk}\Omega_{jk}\eta_j\eta_k + \frac{1}{2}\sum_{\alpha\beta}\chi_{\alpha\beta}\varepsilon_\alpha\varepsilon_\beta \\
&\quad + \sum_{i\alpha j}L_{i\alpha j}u_{i\alpha}\eta_j + \sum_{i\alpha}M_{i\alpha}\eta_i\varepsilon_\alpha + \sum_{i\alpha\beta}N_{i\alpha\beta}u_{i\alpha}\varepsilon_\beta
\end{aligned} \quad (15)$$

where $i$ and $j$ are the atom indices, and $\alpha$ and $\beta$ are the Cartesian indices $\{xyz\}$ for atoms $i$ and $j$. The stain index run over (1 ~ 6) in Voigt notation. The first order expansion parameters are polarization $P_\alpha$, force on the $i$-th atom $F_m$ and stress $\sigma_j$, respectively. Henceforth we shall assume that the atomic coordinates and strains are fully relaxed in the reference system, so that $\frac{\partial E}{\partial u_{i\alpha}} = \frac{\partial E}{\partial \eta_j} = 0$. While the corresponding second-order coefficients are force constant $\Phi_{ij}^{\alpha\beta}$, dielectric susceptibility tensor $\chi_{\alpha\beta}$, ion elastic tensor $\Omega_{jk}$, Born effective charge tensor $N_{i\alpha\beta}$, force-response internal-strain tensor $L_{i\alpha j}$ and frozen-ion piezoelectric tensor $M_{\alpha j}$, respectively. All the second order parameters can be obtained by the density functional perturbation theory (DFPT) calculations.

The equilibrium conditions require that the first derivatives of total energy $E$ regarding to the ionic displacement $u_{i\alpha}$ and strain $\eta_j$ are zero:

$$\begin{aligned}
\frac{\partial E}{\partial u_{i\alpha}} &= \sum_{j\beta}\Phi_{ij}^{\alpha\beta}u_{j\beta} + \sum_j L_{i\alpha j}\eta_j + \sum_j N_{i\alpha\beta}\varepsilon_\beta = 0 \\
\frac{\partial E}{\partial \eta_i} &= \sum_j \Omega_{ij}\eta_j + \sum_{j\alpha}L_{j\alpha i}u_{j\alpha} + \sum_\alpha M_{i\alpha}\varepsilon_\alpha = 0
\end{aligned} \quad (16)$$

Suppose there are $n$ ions in a unit cell. Since the ionic displacement $u_{i\alpha}$ has $3n$ components and the strain $\eta_j$ has 6 components, there are totally $3n + 6$ linear equations. By given a series of small electric field $\varepsilon$ and solving Eq. (16), we can get the relationship between $u$, $\eta$ and $\varepsilon$, then the first derivatives $\frac{\partial u_{i\alpha}}{\partial \varepsilon}$ and $\frac{\partial \eta_i}{\partial \varepsilon}$ can be obtained numerically. The total derivative of $J$ with respect to $\varepsilon$ is:

$$\frac{dJ}{d\boldsymbol{\varepsilon}} = \frac{\partial J}{\partial \boldsymbol{\varepsilon}} + \sum_{i\alpha} \frac{\partial J}{\partial u_{i\alpha}} \frac{\partial u_{i\alpha}}{\partial \boldsymbol{\varepsilon}} + \sum_i \frac{\partial J}{\partial \eta_i} \frac{\partial \eta_i}{\partial \boldsymbol{\varepsilon}} \tag{17}$$

To calculate the $\frac{dJ}{d\varepsilon}$ in Eq. (17), we need calculate three first derivative parameters $\frac{\partial J}{\partial \varepsilon}$, $\frac{\partial J}{\partial u_{i\alpha}}$ and $\frac{\partial J}{\partial \eta_i}$ firstly (see section 3.1 for the calculation details). After the exchange interaction parameter $J$ obtained under a small external electric field $\varepsilon$, the total magnetization $M(\varepsilon)$ can be achieved by performing MC simulation or conjugate gradient (CG) minimization. Hence, the ME coupling parameters can be deduced with $\alpha_{ij} = \frac{\partial M_i(\varepsilon)}{\partial \varepsilon_j}$.

The advantage of our method is that no external field is required, and the ME coupling tensor can be calculated only by calculating the corresponding parameters through the DFT method. Note that the $\frac{dJ}{d\varepsilon}$ consists of three parts, the first part can be regarded as the electron contribution: $\frac{\partial J}{\partial \varepsilon}$, and the other two parts ($\frac{\partial J}{\partial u_{i\alpha}}$ and $\frac{\partial J}{\partial \eta_i}$) are lattice contributions. Therefore, the ME tensor $\alpha$ calculated through this method can also be divided into the contributions of electrons $\alpha^{el}$ and ions $\alpha^{latt}$ for further analysis.

## 3. Computational methods and PASP code

In Sect. 2, we introduce various theoretical models for ferroic systems. In this section, we discuss how to calculate the parameters in these models within the framework of first-principles calculations. The "four-state" method [56] is introduced here, which can be adopted to calculate the isotropic Heisenberg $J$, full $\mathcal{J}$ matrix, SIA $\mathcal{A}$ matrix in magnetic model, and expansion coefficients $\boldsymbol{P}_1^{\alpha\beta}$ and $\boldsymbol{P}_{es}^{12}$ in the unified model, and the parameters $\frac{\partial J}{\partial \varepsilon}$, $\frac{\partial J}{\partial u_{k\alpha}}$, $\frac{\partial J}{\partial \eta_k}$ in ME coupling constant calculations. To construct a more complicated effective Hamiltonian model, a machine learning method was also described. At last, the PASP software which implemented these models and methods will be briefly introduced.

## 3.1 Four-state method

### 3.1.1 Second-order magnetic parameters

For isotropic Heisenberg exchange interaction $J$ (without SOC), it can be deduced through setting the following four spin configurations of $i$-th and $j$-th magnetic ions: (i) spin up states for both $S_i$ and $S_j$: (↑, ↑); (ii) spin up state for $S_i$ and spin down state for $S_j$: (↑, ↓); (iii) spin down state for $S_i$ and spin up state for $S_j$: (↓, ↑); (iv) spin down states for both $S_i$ and $S_j$: (↓, ↓). The other spins are set according to experimental spin state or a low-energy state and remain unchanged in the four spin states. By calculating the total energies, the exchange interaction $J$ (neglect the subscript $ij$ for short) can be determined by $J = \frac{E_1+E_4-E_2-E_3}{4}$. In addition, the bilinear interactions with the effect of SOC can also be calculated by this four-state method in a similar way, such as $\mathcal{J}$ and $\mathcal{A}$ matrices. [55]

### 3.1.2 Parameters in unified polarization model

The calculation of exchange striction interaction parameter $P_{es}^{12}$ is similar with the calculation of Heisenberg $J$. With setting four noncollinear spin arrangements on the spin dimer, the parameter $P_{es}^{12}$ can be derived with mapping polarizations of four state with $P_{es}^{12} = \frac{P_1+P_4-P_2-P_3}{4}$, where $P_i$ is the electric polarization of the $i$-th state. When the SOC effect is considered, the intersite terms $P_{12}^{\alpha\beta}$ and single site terms $P_1^{\alpha\beta}$, $P_1^{\alpha\alpha} - P_1^{zz}$ ($\alpha, \beta = x, y, z$) can be calculated similarly. [73,74]

### 3.1.3 Parameters of linear ME effect

DFT calculations can give atomic forces and stress for a given structure and a given magnetic configuration. The four spin states setting on the $i$-th and $j$-th magnetic ions are same as in the case of the isotropic Heisenberg $J$. Differentiating the exchange parameter $J$ with respect to electric field $\varepsilon$, ionic displacement $u_{k\alpha}$ and strain $\eta_k$, the corresponding parameters of linear ME effect can be obtained with the following expression: $\frac{\partial J}{\partial \varepsilon} = -\frac{P_1+P_4-P_2-P_3}{4}$; $\frac{\partial J}{\partial u_{k\alpha}} = -\frac{F_1+F_4-F_2-F_3}{4}$ and $\frac{\partial J}{\partial \eta_k} = -\frac{\sigma_1+\sigma_4-\sigma_2-\sigma_3}{4}$. [75]

### 3.2 Machine learning method for constructing effective Hamiltonian

In principle, all the magnetic parameters can be derived by the energy mapping method. The four-state method can be used to calculated the bilinear interactions and biquadratic interactions, but it has a limitation for calculating the higher order interactions. However, the higher order interactions can not be neglect in some situations when describing the physical properties, especially in metallic and narrow gap magnets. Thus, a machine learning (ML) approach [89] was developed where the important interaction parameters can be extracted with multiple linear regression analyses and adopting several ML techniques. In our ML method, we first set manually the truncation distance and the highest order of the interactions. Then, we will obtain all possible interactions according to the truncation order and truncation distance by performing group theory analysis. The important interactions are selected automatically with the ML approach. The decision of the truncation distance and order is mainly based on experience and a small amount of testing. For magnetic systems, it is usually enough to consider spin interactions up to the fourth order within the distance of 10 Å. For ferroelectrics, it may be necessary to keep the interactions up to the sixth order.

The ML approach is a general approach to construct the effective Hamiltonian models. For example, the ML approach has been well applied to study the magnetic interactions in the multiferroic material $TbMnO_3$, [90] monolayers $NiX_2$ (X = C, Br, I) [67] and $Fe_3GeTe_2$. [68] Our ML approach can not only reproduce previous results, but also can predict the other important magnetic interactions, such as three-body fourth-order interactions.

### 3.3 Introduction to PASP

In order to study the properties of complex condensed matter systems and overcome the insufficiency of the DFT calculations (e.g., size limit and unable to calculate finite temperature properties), we have developed a software package which is named PASP (Property Analysis and Simulation Package for materials). [90] Almost all models and methods discussed above were implemented in PASP. Based on the first principle calculations, PASP is able to simulate the thermodynamic properties of complex systems and provide insight into the microscopic mechanisms of the coupling between multiple degrees of freedoms.

The PASP package can be used in conjunction with first principle calculation packages, e.g., VASP and Quantum Espresso (QE) etc. Before the DFT calculations, our package can automatically generate a general form of the Hamiltonian according to the symmetry analyses and can generate a set number of configurations with a given supercell. Then these configurations will be prepared for running the first principle calculation. When the DFT calculations done, the parameters in Hamiltonian can be fitted with PASP code and the effective Hamiltonians will be constructed. Based on this effective Hamiltonian, the PASP package can be adopted to perform parallel tempering Monte Carlo (PTMC) simulation [91,92] which can predict the ground state and thermodynamic properties. Our current PASP package includes several functionalities as follows:

(A) **Symmetry and group theory analysis**. The PASP software can identify the symmetry (including point group, space group, and magnetic group) of the system. It can also identify the irreducible representation (IR) of the Bloch wave function, which is convenient for further analysis, such as constructing the $\boldsymbol{k}\cdot\boldsymbol{p}$ Hamiltonian. In addition, we added some modules to facilitate the usage of four-state method (Sec. 3.1). All nonequivalent magnetic pair interactions within a given cutoff radius can be identified automatically. For each magnetic pair, PASP will output the four magnetic configurations for the use in subsequent DFT calculations.

(B) **Global structure searching**. In our package, we implemented the basin-hopping method [93] and genetic algorithms (GA) to predict the structures of clusters, 2D and 3D crystals, and interfaces. Recently, we implement the GA based global optimization approach with explicitly considering the magnetic degree of freedom. [94]

(C) **Effective Hamiltonian methods**. The effective Hamiltonian method can be applied to study the different kinds of physical properties. Correspondingly, several different types of effective Hamiltonian methods are implemented in this module. **(i) Tight binding (TB) Hamiltonian.** The TB model is often used to calculate the electronic states of a material and understand the mechanism of magnetism and ferroelectricity. **(ii) Effective Atomistic Hamiltonian.** In PASP, we implemented the effective Hamiltonian approach [see Sec. 2, Eq. (1)] including the spin Hamiltonian model (Sec. 2.1), and effective Hamiltonian for ferroelectrics (Sec. 2.2). Besides, the unified polarization model is implemented in order to compute the spin order induced polarization (Sec. 2.3). **(iii) Machine learning method for constructing realistic effective Hamiltonian.** Besides the simple four-state method, the ML method discussed in Sec. 3.2 is integrated to PASP to calculate the higher-order magnetic interactions (Sect. 2.1.4). **(iv) Machine learning potential.** Instead of the usual polynomial form, here the artificial neural network (ANN) is used to describe the potential-energy surface of magnets and ferroelectrics.

(D) **Monte Carlo simulation module**. In our package, we implement not only the usual Metropolis MC algorithm but also the parallel-tempering (PT) algorithm, which can be regarded as a parallelized version of the simulated tempering but with different extended ensemble. [95,96] Our effective PTMC method was adopted to predict the ground state and phase transition in different kinds of systems. [50−52,68]

## 4. Applications

In this section, the application of the effective Hamiltonian method in the PASP package to three typical systems (2D FE SnTe, [51] the 2D FM $CrI_3$ and $CrGeTe_3$ [60] and the layered multiferroic $MnI_2$ [73]) will be described to demonstrate the power of the method and software.

### 4.1 2D Ferroelectric SnTe

In the perfectly defect-free SnTe thin films, in order to investigate how $T_c$ intrinsically changes with the layer thickness, an effective Hamiltonian is constructed to estimate its $T_c$. A simple bulk structure is considered where the dipoles form a tetragonal lattice and the FE polarization is along the in-plane [110] direction.

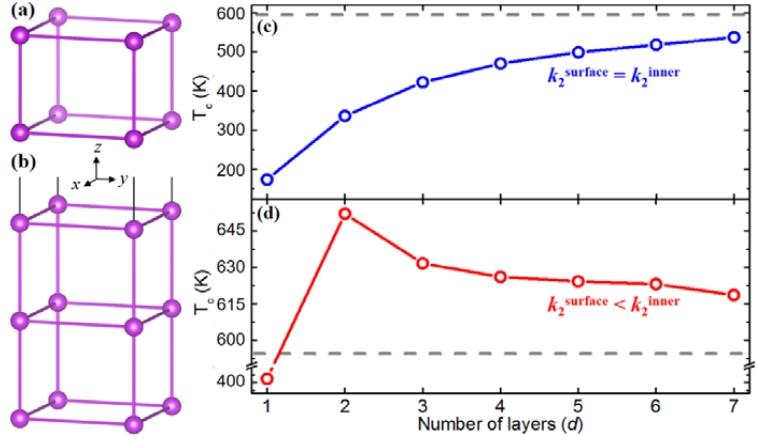

**Fig. 1** $T_c$ as a function of film thickness estimated with the simple effective Hamiltonian. The bulk structure and 3-layer thin film structure are shown in (a) and (b), respectively. The results for $k_2^{sur} = k_2 = 0.05$ and $k_2^{sur} = -0.1 < k_2 = 0.05$ are shown in Panels (c) and (d), respectively. The other parameters are $k_4 = 0.4$ and $J = -0.1$. The horizontal dashed line represents the $T_c$ for the bulk structure. Reproduced with permission from Ref. [50].

For the bulk structure, a simple Hamiltonian can be written as: $E^{tot} = E^{self}(\{u\}) + E^{short}(\{u\})$. Where $E^{self}(\{u\}) = \sum_i [k_2 \boldsymbol{u}^2 + k_4(u_{ix}^4 + u_{iy}^4) + k_4' u_{ix}^2 u_{iy}^2]$ and we set $k_4 = k_4' > 0$ for simplicity. Only the short-range interaction is considered with $E^{short}(\{u\}) = \sum_{\langle ij \rangle} J(u_{ix} u_{jx} + u_{iy} u_{jy})$ and all nearest neighboring interactions are supposed to be the same (both in-plane and out-of-plane). For the [001] thin films [the atomic model of the 3-layer thin film is shown in Fig. 1(b)], the second-order force constant of the surface dipoles can be different with that of the inner dipoles. To be more specific, the self-energy for the thin film can be written as: $E^{self}(\{u\}) = \sum_{i \in sur} k_2^{sur} u^2 + \sum_{i \in inn} k_2^{inn} u^2 + \sum_i [k_4(u_{ix}^4 + u_{iy}^4) + k_4' u_{ix}^2 u_{iy}^2]$. Where $k_2^{inn}$ is the same as that in the bulk structure (i.e., $k_2^{inn} = k_2$). The simple effective Hamiltonian is used to estimate $T_c$ of thin films, and the results indicated $T_c$ is thickness dependent, as shown in Fig. 1(c) and (d). When $k_2^{sur} = k_2$, $T_c$ increases with the film thickness, in agreement with previous beliefs. [50] Interestingly, $T_c$ has a maximum at a certain thickness when $k_2^{sur} < k_2$. These results show that ferroelectric $T_c$ in the defect-free freestanding thin films could be higher than that in bulk.

### 4.2  2D Ferromagnetic CrI$_3$ and CrGeTe$_3$

CrI$_3$ and CrGeTe$_3$ monolayers were discovered to be ferromagnetic.[9–14] The valence state of chromium ion in these two compounds are all +3, with the $3d^3$ configuration and $S$ = 3/2. The FM state is induced by the super exchange interaction between nearest-neighbor Cr ions which are linked by I or Te ligands with the nearly 90° angles. CrI$_3$ has been demonstrated to be well described by the Ising behavior with the spins pointing parallel or antiparallel to the out-of-plane $z$-direction. In contrast, the magnetic anisotropy of CrGeTe$_3$ was determined to be consistent with the Heisenberg behavior, for which the spins can freely rotate and adopt any direction in the three-dimensional space. The Hamiltonians to describe magnetic properties in CrI$_3$ or CrGeTe$_3$ is: $H = H_{ex} + H_{SIA} = \sum_{\langle ij \rangle} \mathbf{S}_i \cdot J_{ij} \cdot \mathbf{S}_j + \sum_i \mathbf{S}_i \cdot A_{ii} \cdot \mathbf{S}_i$, where $J_{ij}$ and $A_{ii}$ are 3 × 3 matrices gathering exchange interaction and SIA parameters, respectively. Three different coordinate systems {$xyz$}, {$XYZ$} and {$\alpha\beta\gamma$} are shown in Fig. 2. In CrI$_3$, the calculated $\mathcal{J}$ matrix is symmetric, and the $J_{xx}$, $J_{yy}$ and $J_{zz}$ are −2.29, −1.93, and −2.23 meV, respectively. The diagonal form $J_\alpha$, $J_\beta$ and $J_\gamma$ are −2.46, −2.41 and −1.59 respectively. The $J_\alpha$ and $J_\beta$ in CrGeTe$_3$ are about −6.65 meV, while $J_\gamma$ = −6.28 meV is about 0.4 meV smaller in magnitude than the other two exchange coefficients.

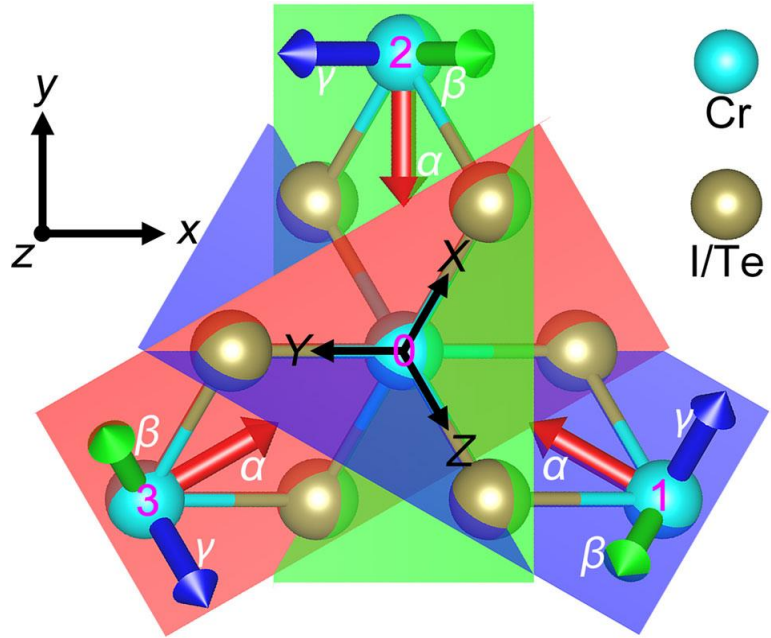

**Fig. 2** Schematization of the CrI$_3$ and CrGeTe$_3$ structures, as well as the different coordinate systems indicated in the text. The planes in blue, green and red indicate the easy plane form Kitaev interaction for Cr$_0$-Cr$_1$, Cr$_0$-

$Cr_2$ and $Cr_0$-$Cr_3$ pairs, respectively. Note that Ge of $CrGeTe_3$ is not shown for simplicity. Reproduced with permission from Ref. [60].

The Kitaev interaction $K = J_\gamma - J_\alpha$ in $CrI_3$ and $CrGeTe_3$ are 0.85 meV and 0.36 meV, respectively; and $A_{zz} = -0.26$ meV in $CrI_3$ and 0.25 meV in $CrGeT_3$. The total Hamiltonian can then be rewritten as: $H = \sum_{\langle ij \rangle}(J\boldsymbol{S}_i \cdot \boldsymbol{S}_j + KS_i^\gamma S_j^\gamma) + \sum_i A_{zz}(S_i^z)^2$. In the case of $CrI_3$, both Kitaev interaction and SIA contribute to the out-of-plane easy axis behavior, therefore explaining the Ising behavior of $CrI_3$. In the case of $CrGeTe_3$, Kitaev interaction competes with SIA, and $\Delta\varepsilon$ is nearly zero ($-0.01$ meV/f.u.), which implies isotropic magnetic behavior in $CrGeTe_3$.

### 4.3 Layered Multiferroics $MnI_2$

The magnetic ground state of layered $MnI_2$ is proper screw magnetic configuration and no polarization occurs in zero external magnetic field. Under a low in-plane magnetic field ($H <$ 3T), the proper screw spin order adopts the in-plane component $\boldsymbol{q} \parallel [1\bar{1}0]$, with the polarization $\boldsymbol{P} \parallel [110] \perp \boldsymbol{q}$. When the external in-plane magnetic field is in the range of 3–7 T, the polarization $\boldsymbol{P}$ is parallel to $\boldsymbol{q} \parallel [110]$. In both cases the polarization can flop when rotating the applied $H$ along the [001] direction, which implies the strong ME coupling in $MnI_2$.

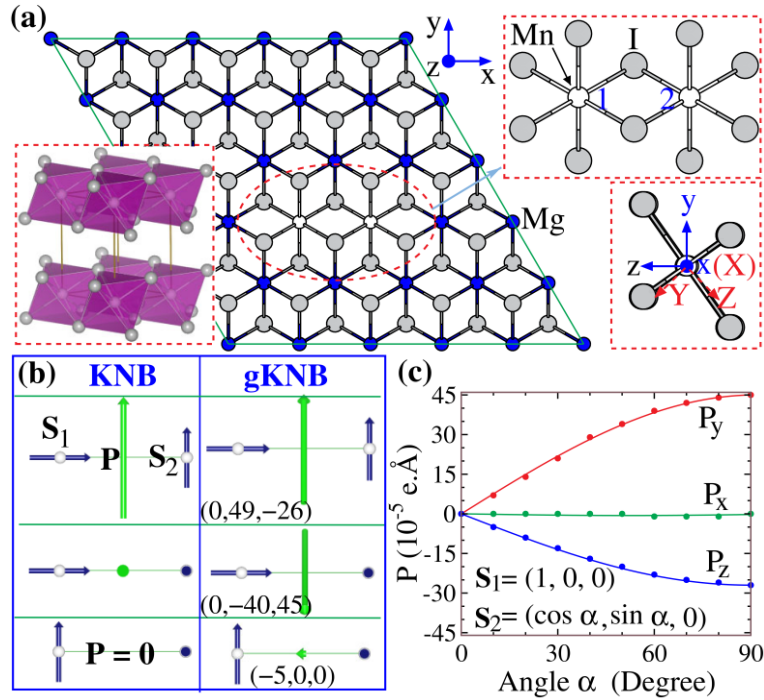

**Fig. 3** (a) The 5 × 5 × 1 supercell of MnI2. The left inset illustrates the layered structure of MnI2. (b) The electric polarizations predicted by the KNB and gKNB models for three different spin configurations of the Mn-Mn dimer. (c) The polarization of the Mn-Mn pairs extracted by the gKNB model (line) and direct density functional calculation (dot). Reproduced with permission from Ref. [73].

The multiferroicity in MnI2 can not be explained by any of the previous models. The unified model was adopted to study the polarization in MnI2. [73] Note that spatial inversion symmetry exists at the center of the Mn-Mn spin dimer and the fact that $Mn^{2+}$ ion itself is an inversion center, the intrasite polarization $P_i^{\alpha\beta}$ and the diagonal coefficients of the intersite polarization should vanish. The values in $M$ matrix of the intersite polarization are calculated by four-state method. Our results differ from the KNB model [72,73] because the matrix elements $M_{11} = (P_{12}^{yz})_x$, $M_{22} = (P_{12}^{zx})_y$ and $M_{11} = (P_{12}^{xy})_z$ are non-zero and $M_{23} = (P_{12}^{yz})_x \neq -M_{23} = (P_{12}^{zx})_z$. In the helical spin-spiral order, the sum of the intrasite polarization is zero, while the intersite polarization for each site is $P_0^{tot} = \sum_{k=1}^{6} P_{0k} = \sum_{k=1}^{6} M^{0k}(S_0 \times S_k)$, and the total polarization for site 0 is shown in Fig. 3(b). For $q = (Q, 0, 0)$, $P_0^{tot} = \left(\frac{\sqrt{3}}{2}A, -\frac{3}{2}A, 0\right)$. In the case of $q = (Q, Q, 0)$, $P_0^{tot} = \left(-\frac{1}{2}B, \frac{\sqrt{3}}{2}B, 0\right)$. Thus, our gKNB model predicts that $P \perp q$ in the low applied field with $q = (Q, 0, 0)$, but $P \parallel q$ in the high applied field with $q = (Q, Q, 0)$, consistent with the experiment.

## 5. Summary and perspective

In this review, we summarize the recent theoretical and computational advances in magnetics, ferroelectrics and multiferroics. We first discuss the effective Hamiltonian of magnetics and ferroelectrics, and then describe the unified polarization model for multiferroics and the method for calculating the ME coupling tensor in details. Next, we discussed the "four-state" method within the DFT framework, and the PASP package, and present three examples to illustrate the typical applications of the PASP package.

Although first principle based approaches have made important progress in understanding the mechanisms of magnetism and ferroelectricity, there are still big challenges to be overcome when simulating realistic materials and predicting new materials. For example, the magnetic properties depend on the adopted Hubbard U parameter which is hard to estimate accurately; local density approximation (LDA) will usually underestimate the ferroelectricity, while the widely adopted generalized gradient approximation Perdew-Burke-Ernzerhof (PBE) functional will overestimate the ferroelectricity. DFT calculations cannot directly simulate the flipping dynamics of ferroelectric, magnetic and multiferroic domains under external fields. These issues can be solved by developing new DFT functionals or developing more efficient and more accurate effective Hamiltonian approaches.

Besides the limits of current DFT in studying ferroic materials to be overcome, in our opinion some other research directions in this field remains to be explored: (I) Predicting and designing high-performance (especially high critical temperature, low dimensional) ferroic materials is still an important topic; (II) Rare earth compounds may host exotic states (e.g., quantum spin liquid, multiferroicity, or heavy-fermions), while reliable model Hamiltonians applicable to the f-electron (lanthanides and actinides) systems are lacking; (III) The interaction between strong optical fields and ferroic materials may give rise to unusual phenomena, which deserve much more research attention.

## Acknowledgements

This work are supported by NSFC (Grant Nos. 11825403, 12188101, and 11804138), Anhui Provincial Natural Science Foundation (1908085MA10), and the Opening Foundation of State Key Laboratory of Surface Physics Fudan University (Grant No. KF2019_07). We thank Dr. Zhang H M for useful discussions.

## References

[1]  Li X Y, Yu H Y, Lou F, Feng J S, Whangbo M H and Xiang H J 2021 *Molecules* **26** 803

[2]  Binder K and Young A P 1986 *Rev. Mod. Phys.* **58** 801–976

[3]  Castelnovo C, Moessner R and Sondhi S L 2008 *Nature* **451** 42–45

[4]  Yan S, Huse D A and White S R 2011 *Science* **332** 1173–1176

[5] Han T H, Helton J S, Chu S, Nocera D G, Rodriguez-Rivera J A, Broholm C and Lee Y S 2012 *Nature* **492** 406–410

[6] Muehlbauer S, Binz B, Jonietz F, Pfleiderer C, Rosch A, Neubauer A, Georgii R and Boeni P 2009 *Science* **323** 915–919

[7] Neubauer A, Pfleiderer C, Binz B, Rosch A, Ritz R, Niklowitz P G and Boeni P 2009 *Phys. Rev. Lett.* **102** 186602

[8] Fujishiro Y, Kanazawa N and Tokura Y 2020 *Appl. Phys. Lett.* **116** 090501

[9] Gong C, Li L, Li Z L, Ji H W, Stern A, Xia Y, Cao T, Bao W, Wang C Z, Wang Y, Qiu Z Q, Cava R J, Louie S G, Xia J and Zhang X 2017 *Nature* **546** 265–269

[10] Huang B, Clark G, Navarro-Moratalla E, R. Klein D R, Cheng R, Seyler K L, Zhong D, Schmidgall E, McGuire M A, Cobden D H, Yao W, Xiao D, Jarillo-Herrero P and Xu X D 2017 *Nature* **547** 270−273

[11] Song T C, Cai X H, Tu M W Y, Zhang X O, Huang B, Wilson N P, Seyler K L, Zhu L, Taniguchi T, Watanabe K, McGuire M A, Cobden D H, Xiao D, Yao W and Xu X D 2018 *Science* **360** 1214−1218

[12] Seyler K L, Zhong D, Klein D R, Gao S Y, Zhang X O, Huang B, Navarro-Moratalla E, Yang L, Cobden D H, McGuire M A, Yao W, Xiao D, Jarillo-Herrero P and Xu X D 2018 *Nat. Phys.* **14** 277−281

[13] Cardoso C, Soriano D, García-Martínez N A and Fernández-Rossier J 2018 *Phys. Rev. Lett.* **121** 067701

[14] Jiang S, Shan J and Mak K F 2018 *Nat. Mater.* **17** 406−410

[15] Deng Y J, Yu Y J, Song Y C, Zhang J Z, Wang N Z, Sun Z Y, Yi Y F, Wu Y Z, Wu S W, Zhu J Y, Wang J, Chen X H and Zhang Y B 2018 *Nature* **563** 94−99

[16] Bonilla M, Kolekar S, Ma Y J, Diaz H C, Kalappattil V, Das R, Eggers T, Gutierrez H R, Phan M H and Batzill M 2018 *Nat. Nanotechnol.* **13** 289−293

[17] O'Hara D J, Zhu T C, Trout A H, Ahmed A S, Luo Y Q K, Lee C H, Brenner M R, Rajan S, Gupta J A, McComb D W and Kawakami R K 2018 *Nano. Lett.* **18** 3125−3131

[18] Zheng S, Huang C X, Yu T, Xu M L, Zhang S T, Xu H Y, Liu Y C, Kan E J, Wang Y C and Guochun Yang G C 2019 *J. Phys. Chem. Lett.* **10** 2733−2738

[19] Tian S J, Zhang J F, Li C H, Ying T P, Li S Y, Zhang X, Liu K and Lei H C 2019 *J. Am. Chem. Soc.* **141** 5326−5333

[20] Su X Y, Qin H L, Yan Z B, Zhong D Y and Guo D H 2020 *Chin. Phys. B* **31** 037301

[21] Wu M H 2021 *Acs Nano* **15** 9229−9237

[22] Gao W X, Zhu Y, Wang Y J, Yuan G L and Liu J M 2020 *J. Materiomics* **6** 1−16

[23] Xu W J, Romanyuk K, Martinho J M G, Zeng Y, Zhang X W, Ushakov A, Vladimir Shur V, Zhang W X, Chen X M, Kholkin A, and Rocha J 2020 *J. Am. Chem. Soc.* **142** 16990–16998

[24] Li P F, Liao Q, Tang Y, Qiao W C, Zhao D W, Yong Ai Y, Yao Y F and Xiong R G 2019 *Proc. Natl. Acad. Sci. U.S.A.* **116** 5878−5885

[25] Lavrentovich O D 2020 *Proc. Natl. Acad. Sci. U.S.A.* **117** 14629−14631

[26] Prateek, Thakur V K and Gupta R K 2016 *Chem. Rev.* **116** 4260–4317

[27] Yang Q, Wu M H and Li J 2018 *J. Phys. Chem. Lett.* **9** 7160−7163

[28] Xiong F, Zhang X, Lin Z and Chen Y 2018 *J. Materiomics* **4** 139

[29] Belianinov A, He Q, Dziaugys A, Maksymovych P, Eliseev E, Borisevich A, Morozovska A, Banys J, Vysochanskii Y and Kalinin S V 2015 *Nano Lett.* **15** 3808

[30] You L, Liu F C, Li H S, Hu Y Z, Zhou S, Chang L, Zhou Y, Fu Q D, Yuan G L, Dong S, Fan H J, Gruverman A, Liu Z and Wang J L 2018 *Adv. Mater.* **30** 1803249

[31] Collins J L, Wang C, Tadich A, Yin Y, Zheng C, Hellerstedt J, Grubišić-Čabo A, Tang S, Mo S, Riley J, Huwald E, Medhekar N V, Fuhrer M S and Edmonds M T, 2020 *ACS Appl. Electron. Mater.* **2** 213

[32] Yasuda K, Wang X R, Watanabe K, Taniguchi T and Jarillo-Herrero P 2021 *Science* **372** 1458–1462

[33] Fiebig M, Lottermoser T, Meier D and Trassin M 2016 *Nat. Rev. Mater.* **1** 16046

[34] Wang J, Neaton J B, Zheng H, Nagarajan V, Ogale S B, Liu B, Viehland D, Vaithyanathan V, Schlom D G, Waghmare U V, Spaldin N A, Rabe K M, Wuttig M and Ramesh R 2003 *Science* **299** 1719−1722

[35] Kimura T, Goto T, Shintani H, Ishizaka K, Arima T and Tokura Y 2003 *Nature* **426** 55–58

[36] Khomskii D 2009 *Physics* **2** 20

[37] Wang H and Qian X F 2017 *2D Mater.* **4** 015042

[38] Shen W, Pan Y H, Shen S N, Li H, Nie S Y and Mei J 2021 *Chin. Phys. B* **30** 117503

[39] Huang C X, Du Y P, Wu H P, Xiang H J, Deng K M and Kan E J 2018 *Phys. Rev. Lett.* **120** 147601

[40] Luo W, Xu K and Xiang H J 2017 *Phys. Rev. B* **96** 235415

[41] Song Q, Occhialini C A, Ergeçen E, Ilyas B, Amoroso D, Barone P, Kapeghian J, Watanabe K, Taniguchi T, Botana A S, Picozzi S, Gedik N and Comin R 2022 *Nature* **602** 601–605

[42] Bayaraa T, Xu C S, Yang Y L, Xiang H J and Bellaiche L 2020 *Phys. Rev. Lett.* **125** 067602

[43] Amoroso D, Barone P and Silvia Picozzi 2020 *Nat. Commun.* **11** 5784

[44] Nahas Y, Prokhorenko S, Kornev I and Bellaiche L 2017 *Phys. Rev. Lett.* **119** 117601

[45] Xu C S, Nahas Y, Prokhorenko S, Xiang H J and Bellaiche L 2020 *Phys. Rev. B* **101** 241402(R)

[46] Fujiyama S and Kato R 2019 *Phys. Rev. Lett.* **122** 147204

[47] Leiner J C, Jeschke H O, Valentí R, Zhang S, Savici A T, Lin J Y Y, Stone M B, Lumsden M D, Hong J W, Delaire O, Bao W and Broholm C L 2019 *Phys. Rev. X* **9** 011035

[48] Wu X F, Li J Y, Ma X M etal 2020 *Phys. Rev. X* **10** 031013

[49] Xu B, Dupé B, Xu C S, Xiang H J and Bellaiche L 2018 *Phys. Rev. B* **98** 184420

[50] Liu K, Lu J L, Picozzi S, Bellaiche L and Xiang H J 2018 *Phys. Rev. Lett.* **121** 027601

[51] Gu T, Scarbrough T, Yang Y R, Íñiguez J, Bellaiche L and Xiang H J 2018 *Phys. Rev. Lett.* **120** 197602

[52] Wang D W, Liu L J, Liu J, Zhang N and Wei X Y 2018 *Chin. Phys. B* **27** 127702

[53] Moriya T 1960 *Phys. Rev* **120** 91−98

[54] Hoffmann M and Blügel S 2020 *Phys. Rev. B* **101** 024418

[55] Xiang H J, Kan E J, Wei S H, Whangbo M H and Gong X G 2011 *Phys. Rev. B* **84** 224429

[56] Xiang H J, Lee C H, Koo H J, Gong X G and Whangbo M H 2013 *Dalton Trans.* **42** 823

[57] Kitaev A 2006 *Ann. Phys.* **321** 2–111

[58] Banerjee A, Yan J Q, Knolle J, Bridges C A, Stone M B, Lumsden M D, Mandrus D G, Tennant D A, Moessner R and Nagler S E 2017 *Science* **356** 1055–1059

[59] Takagi H, Takayama T, Jackeli G, Khaliullin G and Nagler S E 2019 *Nat. Rev. Phys.* **1** 264–280

[60] Xu C S, Feng J S, Xiang H J and Bellaiche L 2018 *npj Comput. Mater.* **4** 57

[61] Stavropoulos P P, Pereira D and Kee H Y 2019 *Phys. Rev. Lett.* **123** 037203

[62] Xu C S, Junsheng Feng J S, Kawamura M, Yamaji Y, Nahas Y, Prokhorenko S, Qi Y, Xiang H J and Bellaiche L 2020 *Phys. Rev. Lett.* **124** 087205

[63] Kartsev A, Augustin M, Evans R F L, Novoselov K S and Santos E J G 2020 *npj Comput. Mater.* **6** 150

[64] Slonczewski, J. C. 1991 *Phys. Rev. Lett.* **67** 3172–3175

[65] Lorenz B, Wang Y Q and Chu C W 2007 *Phys. Rev. B* **76** 104405

[66] Fedorova N S, Bortis A, Findler C and Spaldin N A 2018 *Phys. Rev. B* **98** 235113

[67] Ni J Y, Li X Y, Amoroso D, He X, Feng J S, Kan E J, Picozzi S and Xiang H J 2021 *Phys. Rev. Lett.* **127** 247204

[68] Xu C S, Li X Y, Chen P, Zhang Y, Xiang H J and Bellaiche L 2022 *Adv. Mater.* **34** 2107779

[69] Paul S, Haldar S, von Malottki S and Heinze S 2020 *Nat. Commun.* **11** 4756

[70] Zhong W, Vanderbilt D and Rabe K M 1995 *Phys. Rev. B* **52** 6301−6312

[71] Íñiguez J and Vanderbilt D 2002 *Phys. Rev. Lett.* **89** 115503

[72] Katsura H, Nagaosa N and Balatsky A V 2005 *Phys. Rev. Lett.* **95** 057205

[73] Xiang H J, Kan E J, Zhang Y, Whangbo M H and Gong X G 2011 *Phys. Rev. Lett.* **107** 157202

[74] Xiang H J, Wang P S, Whangbo M H and Gong X G 2013 *Phys. Rev. B* **88** 054404

[75] Lu X Z, Wu X F and Xiang H J 2015 *Phys. Rev. B* **91** 100405(R)

[76] Wang P S, Lu X Z, Gong X G and Xiang H J 2016 *Comput. Mater. Sci.* **112** 448–458

[77] Feng J S and Xiang H J 2016 *Phys. Rev. B* **93** 174416

[78] Wu X, Vanderbilt D and Hamann D R 2005 *Phys. Rev. B* **72** 035105

[79] Swartz C W and Wu X 2012 *Phys. Rev. B* **85** 054102

[80] Li J, Feng J S, Wang P S, Kan E J and Xiang H J 2021 *Sci. China Phys. Mech.* **64** 286811

[81] He G M, Zhang H M, Ni J Y, Liu B Y, Xu C S and Xiang H J 2022 *Chin. Phys. Lett.* **39** 067501

[82] Fiebig M 2005 *J. Phys. D: Appl. Phys.* **38** R123

[83] Bousquet E, Spaldin N A and Delaney K T 2011 *Phys. Rev. Lett.* **106** 107202

[84] Ricci F and Bousquet E 2016 *Phys. Rev. Lett.* **116** 227601

[85] Garcia-Castro A C, Romero A H and Bousquet E 2016 *Phys. Rev. Lett.* **116** 117202

[86] Xu K, Liang G J, Feng Y, Xiang H J and Feng J S 2020 *Phys. Rev. B* **102** 224416

[87] Dasa T R, Hao L, Liu J and Xu H X 2019 *J. Mater. Chem. C* **7** 13294

[88] Pi M C, Xu X F, He M Q and Chai Y S 2022 *Phys. Rev. B* **105** L020407

[89] Li X Y, Lou F, Gong X G and Xiang H J 2020 *New J. Phys.* **22** 053036

[90] Lou F, Li X Y, Ji J Y, Yu H Y, Feng J S, Gong X G and Xiang H J 2021 *J. Chem. Phys.* **154** 114103

[91] Liu P, Kim B C, Friesner R A and Berne B J 2005 *Proc. Natl. Acad. Sci. U.S.A.* **102** 13749–13754

[92] Earlab D J and Deem M W 2005 *Phys. Chem. Chem. Phys.* **7** 3910–391

[93] Wales D J and Doye J P K 1997 *J. Phys. Chem. A* **101** 5111–5116

[94] Lou F, Luo W, Feng J S and H. Xiang H J 2019 *Phys. Rev. B* **99** 205104

[95] Hukushima K and Nemoto K 1996 *J. Phys. Soc. Jpn.* **65** 1604–1608

[96] Hansmann U H E 1997 *Chem. Phys. Lett.* **281** 140–1